\documentclass[a4paper,11pt]{article}
\usepackage{jcappub} 
\usepackage{lineno}

\def\ba{\begin{eqnarray}}
\def\ea{\end{eqnarray}}

\usepackage{color}

\title{
\vspace*{-4ex}
\begin{flushright}
\hfill {\small LFTC-26-01/110}
\end{flushright}
\vspace{2ex}
{\boldmath Octet baryon electroweak form factors in dense nuclear matter}
}

\author[a]{G.~Ramalho,}
\author[b]{K.~Tsushima,}
\author[a]{and Myung-Ki Cheoun}
\affiliation[a]{Department of Physics and OMEG Institute, Soongsil University, \\
  Seoul 06978, Republic of Korea}
\affiliation[b]{Laborat\'orio de 
  F\'{i}sica Te\'orica e Computacional -- LFTC, \\
  Programa de P\'osgradua\c{c}\~ao em F\'{i}sica Computacional, 
Universidade Cidade de  S\~ao Paulo,  \\
01506-000,  S\~ao Paulo, SP, Brazil}

\emailAdd{gilberto.ramalho2013@gmail.com}

\abstract{
  Motivated by the necessity of developing theoretical models for studying the electroweak structure of baryons in a nuclear medium, we apply a covariant quark model to study interactions of baryons with nuclear matter. 
  The electromagnetic and axial form factors of the octet baryons are determined by combining a covariant quark model that takes into account the meson cloud dressing of the baryon cores, developed for free space, with the quark-meson coupling model in the extension to the nuclear medium.
  We discuss the medium modifications on the electroweak form factors of octet baryons for the range  of densities from $\rho=0$ up to $\rho=2 \rho_0$, where $\rho_0= 0.15$ fm$^{-3}$ is the normal nuclear matter density.
  We also study how the shape of the form factors is modified in finite nuclei due to the profile of the nuclear density distributions compared with calculations using the average density of the nucleus.}

\begin{document}
\maketitle
\flushbottom

\section{Introduction}

Although experimental information about the electroweak structure of the baryons and mesons in nuclear medium is limited, there is evidence that the electromagnetic and axial form factors
of the nucleons are modified in a nuclear medium~\cite{Dieterich01,Strauch03,Gysbers19a,QMCReview,Medium1,Medium2}.
Theoretical models are then fundamental for the understanding of environments with dense nuclear matter, from high energy nucleus-nucleus collisions to the cores of compact stars~\cite{QMCReview,Medium1,Medium2}.

We propose to study the electroweak structure of the baryons in a nuclear medium combining a covariant quark model defined in terms of valence quarks and meson cloud dressing in free space, with a model developed to study the properties of the hadrons in a nuclear medium: the quark-meson coupling (QMC) model~\cite{QMCReview,QMCEMFFMedium5,Lu98a,Lu99a}.

We consider in particular the covariant spectator quark model~\cite{Nucleon,Omega,NSTAR2017}, developed for the study of nucleon excitations in the spacelike region~\cite{PPNP2024,NSTAR}.
The formalism has been applied to the study of the electromagnetic structure of nucleon resonances in the spacelike region~\cite{NSTAR2017,nstars1,nstars1a,nstars2,nstars3,nstars4}, and the timelike region~\cite{Dalitz1,Dalitz2,OctetDecuplet3}, to the electromagnetic structure of the octet and decuplet baryons~\cite{Omega,Octet1,Octet2,OctetDecuplet1,OctetDecuplet2,OctetDecuplet3}, and other baryons~\cite{Baryons1,Baryons2}.
The model has also been used for the study of baryon form factors in the timelike region~\cite{Hyperons12}, for the study of the axial structure of baryons~\cite{Axial} and for the nucleon deep inelastic scattering~\cite{NucleonDIS}.

The extension of the model to the nuclear medium is performed taking into account the modifications of the properties of the hadrons (effective masses and coupling constants) in terms of the nuclear density ~\cite{Medium1,Medium2,Medium3,Medium4} as determined by the QMC model.
The in-medium coupling constants are determined in terms of the axial couplings, baryon masses and pion decay constant $f_\pi$~\cite{Medium1,Medium2}.
To represent the variables in medium: masses, couplings constants, and form factors, we use the upper-script $*$.


\section{Results}

Our formalism has been applied to the electromagnetic and axial structure of the baryon octet for constant densities $\rho=0$ (free space) and $\rho=0.5 \rho_0$, $\rho_0$~\cite{Medium2,Medium3}.
The calculations for the nucleon electroweak form factors can be used to calculate the cross sections associated with the neutrino-nucleus scattering for neutral currents and charged currents~\cite{Cheoun13a,Cheoun13b}.
The conclusion is that the cross sections are, in general, suppressed in a nuclear medium~\cite{Medium3}.

\begin{figure*}[t]
  \vspace{.1cm}
\centerline{
\mbox{
\includegraphics[width=1.9in]{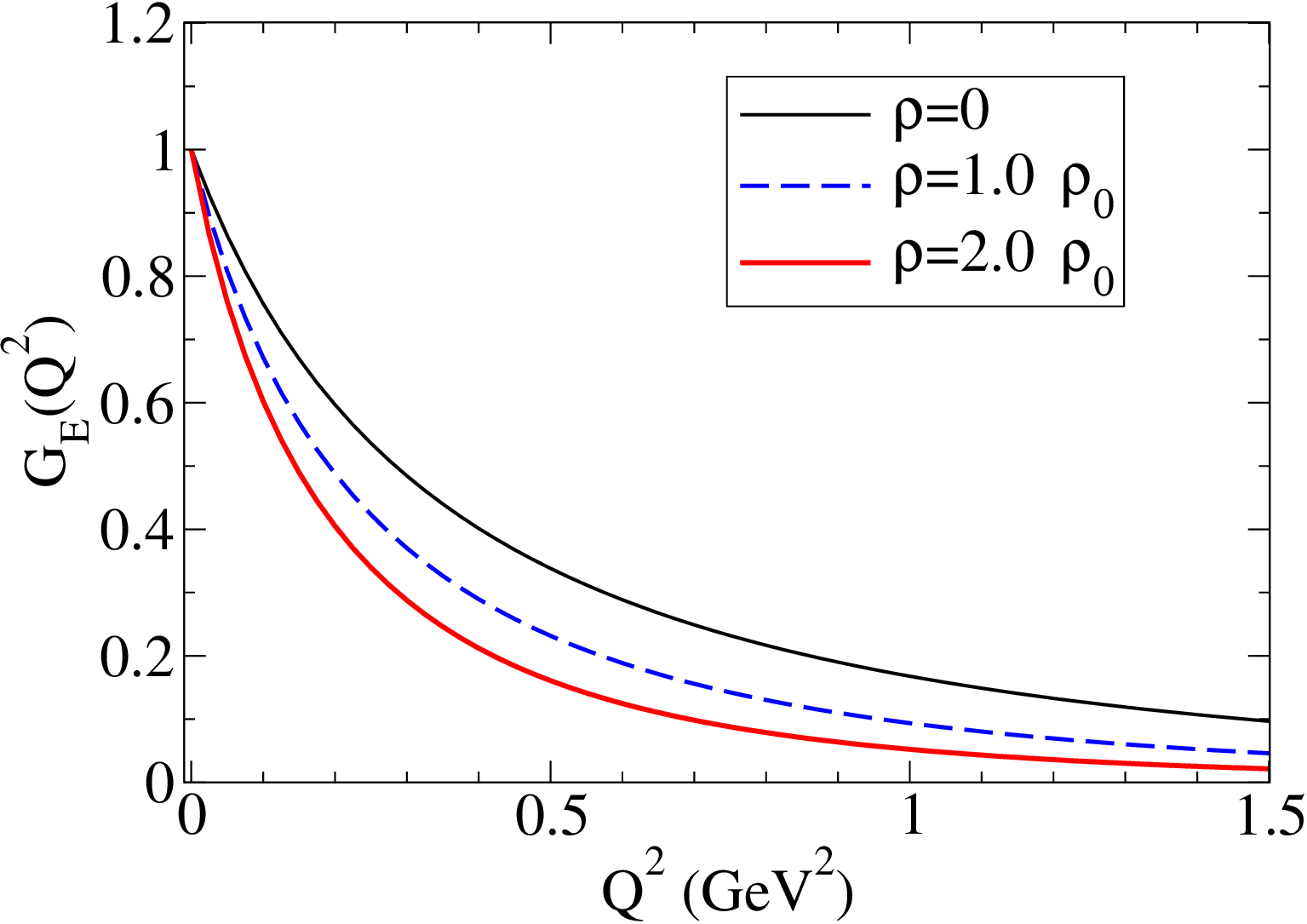} \hspace{.1cm}
\includegraphics[width=1.9in]{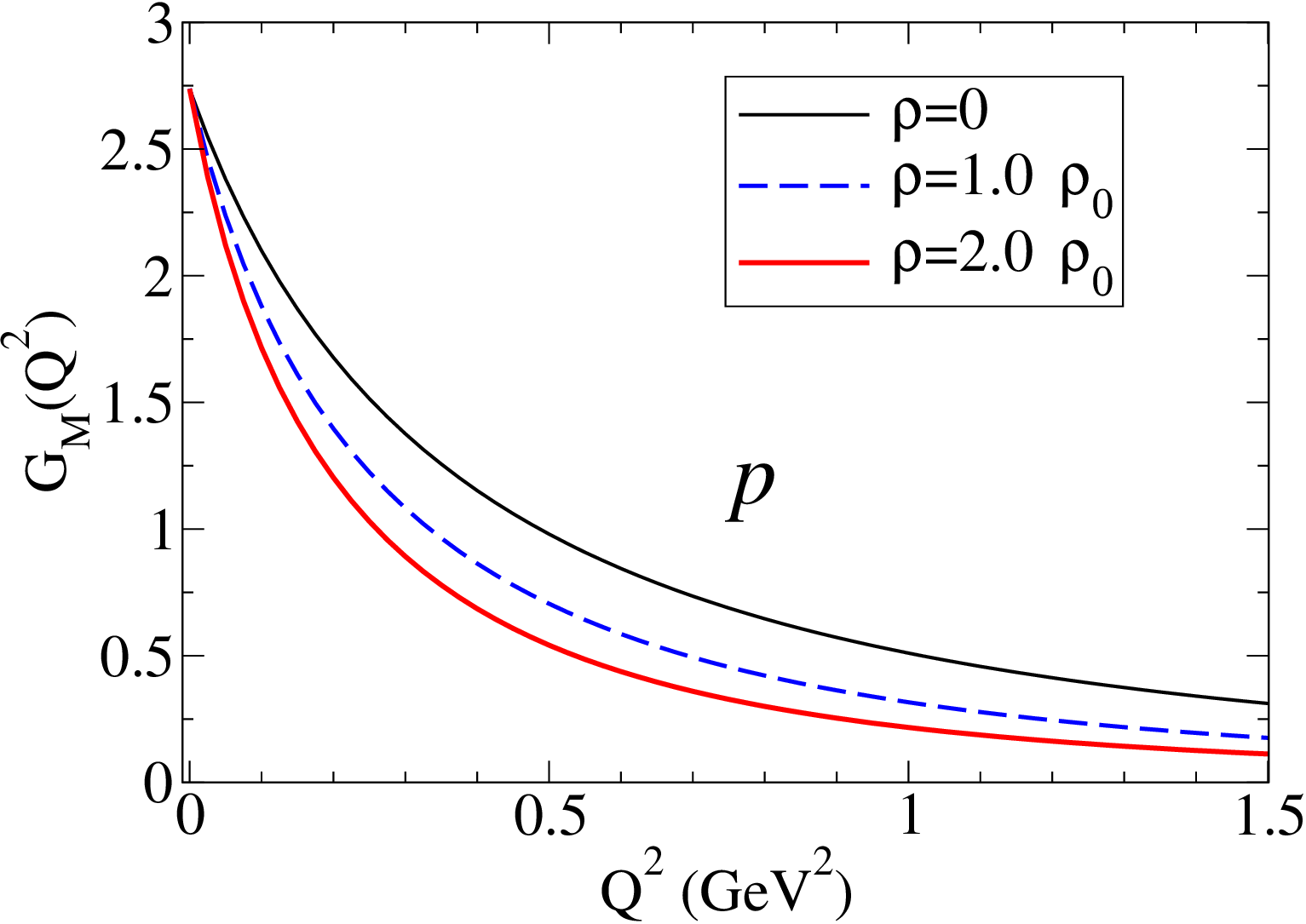} \hspace{.1cm}
\includegraphics[width=1.9in]{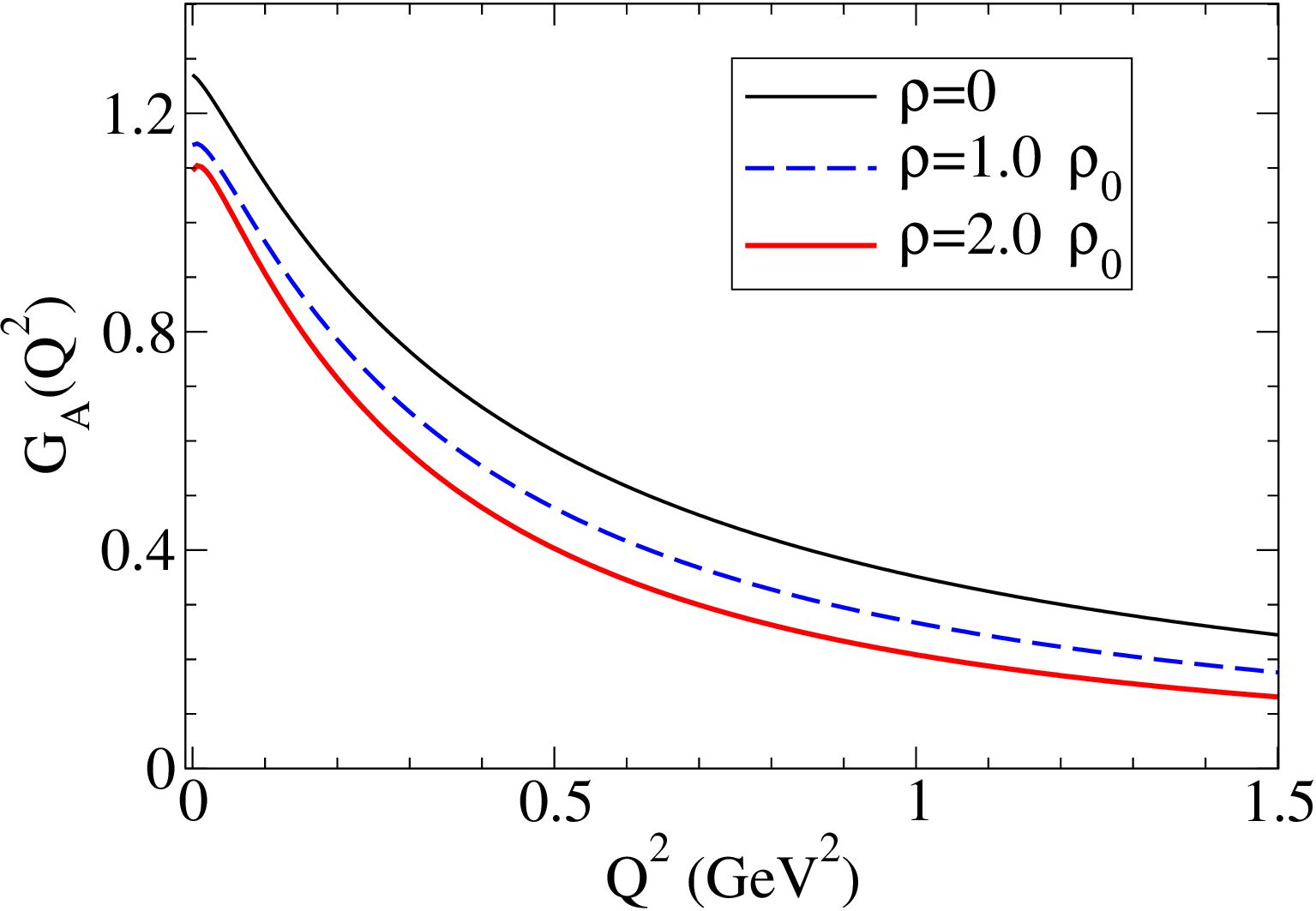}
}}
\vspace{.15cm}
\centerline{
\mbox{
\includegraphics[width=1.9in]{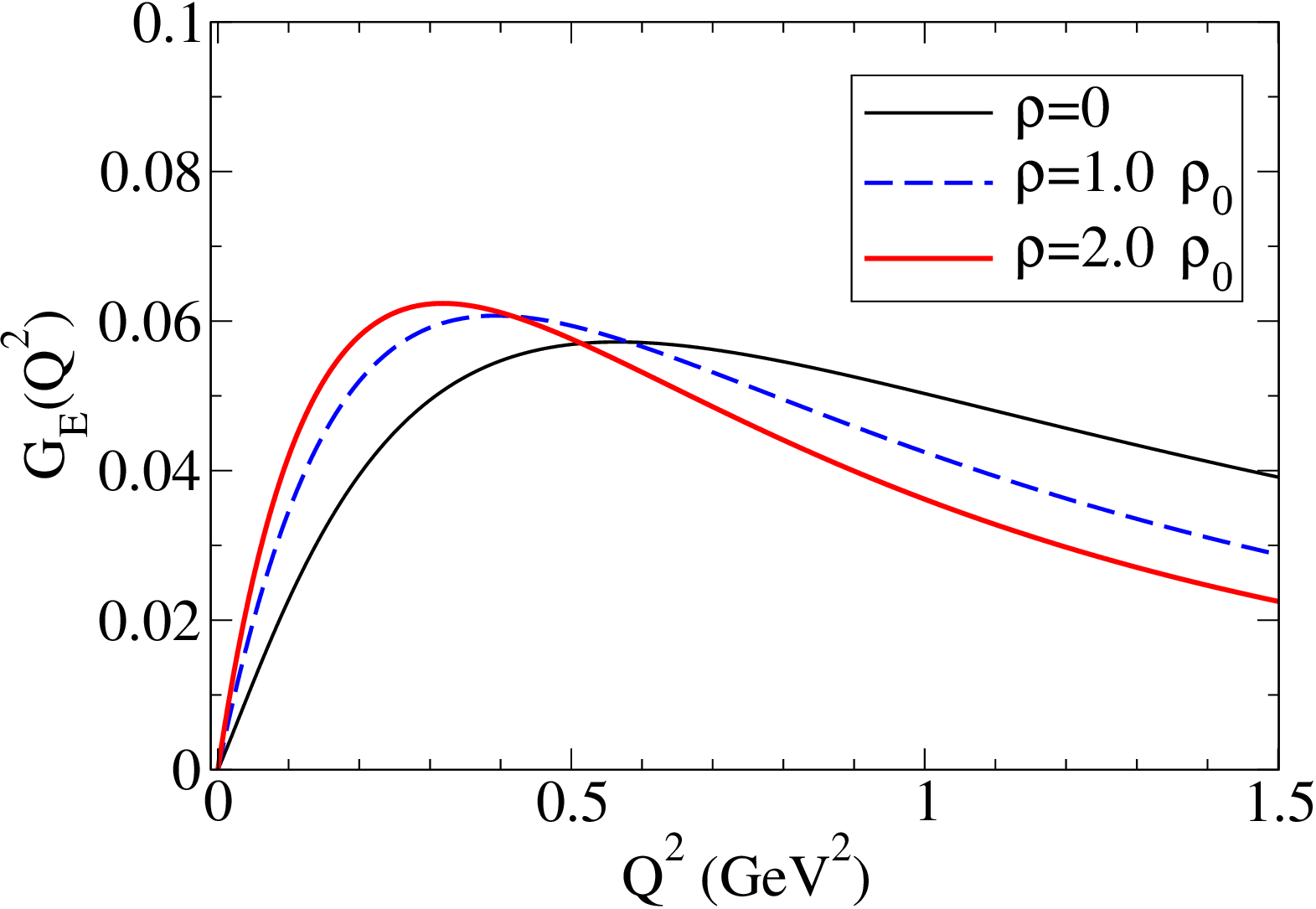} \hspace{.1cm}
\includegraphics[width=1.9in]{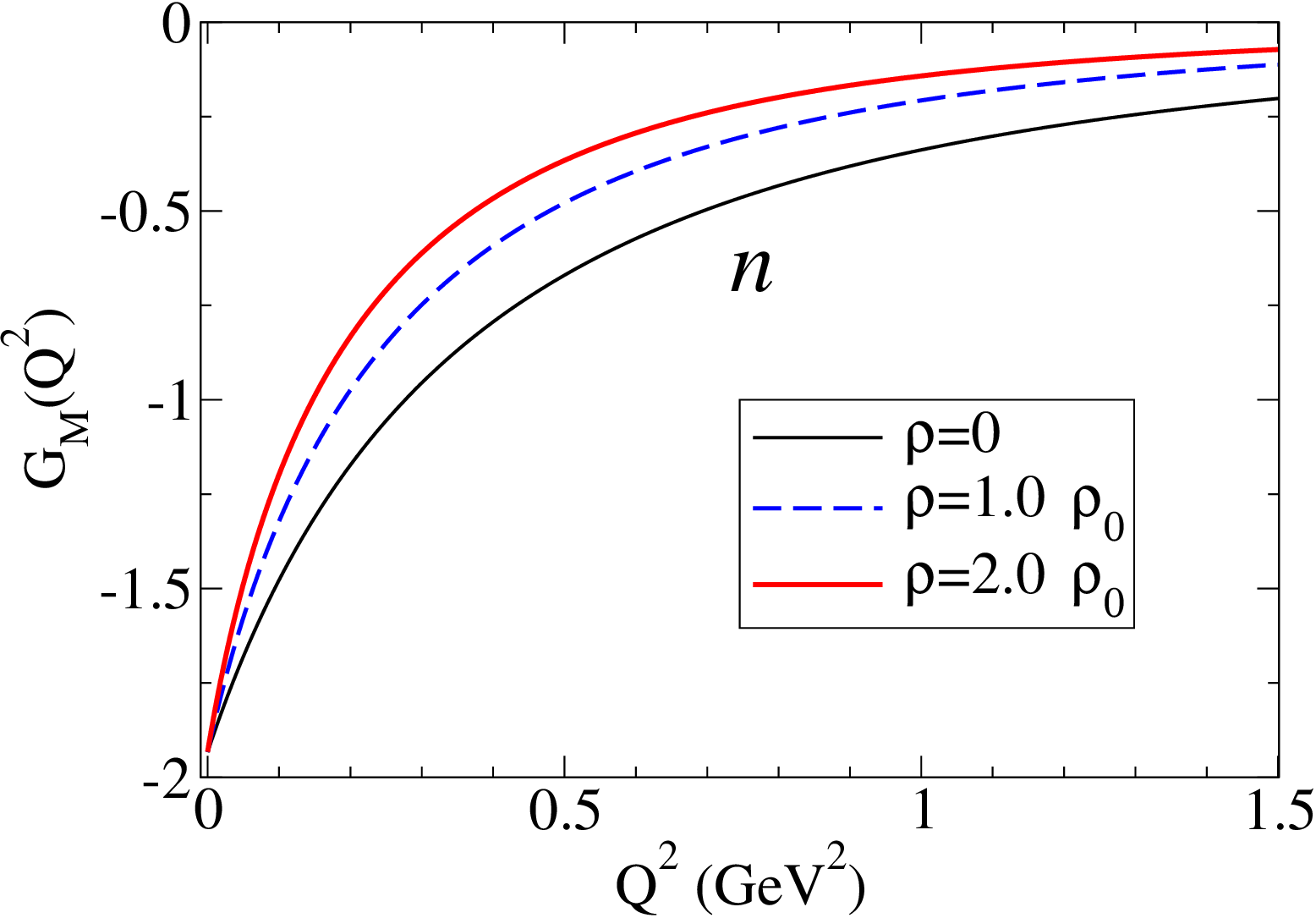} \hspace{.1cm}
\includegraphics[width=1.9in]{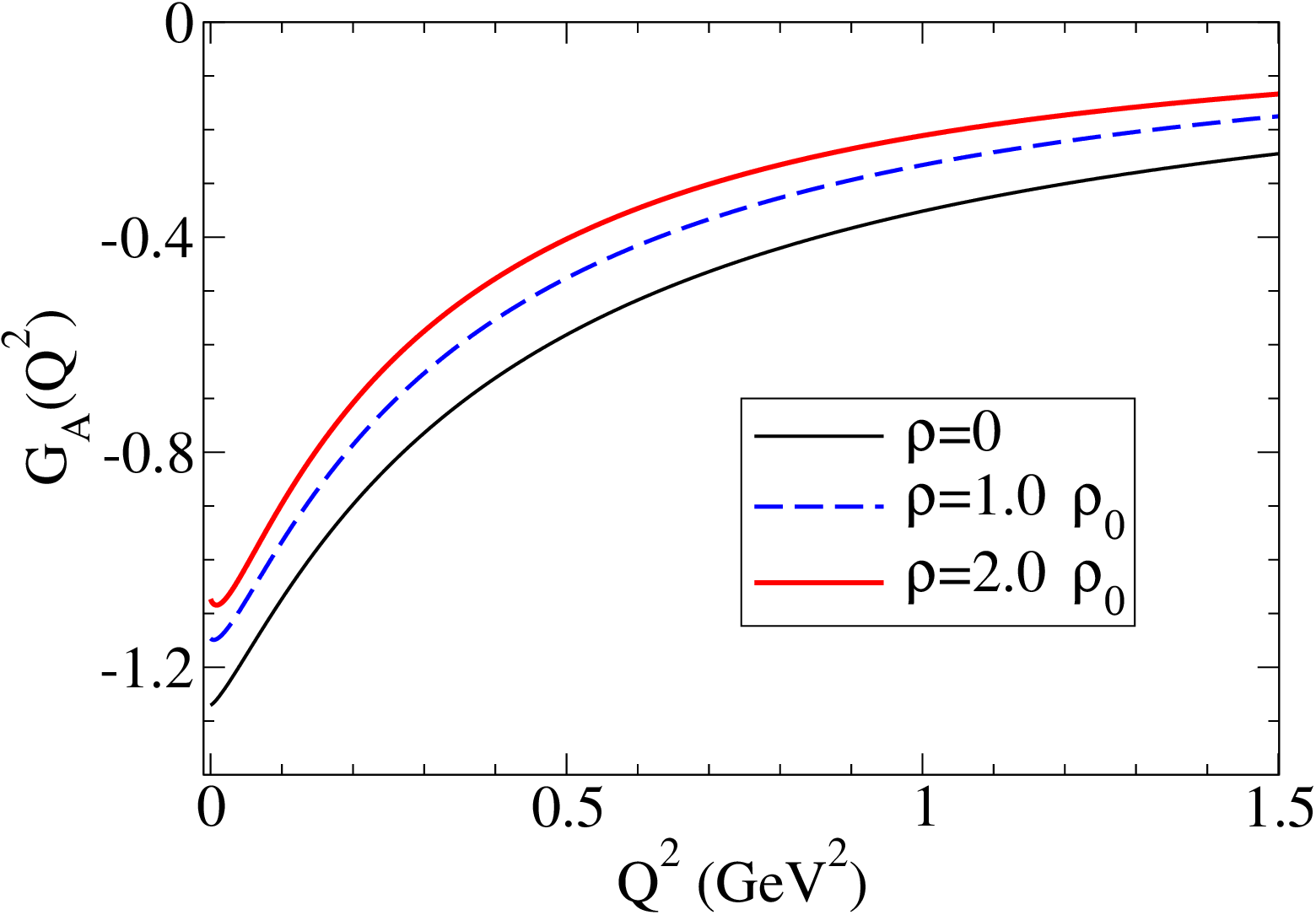}
}}
\caption{\footnotesize{Proton and neutron form factors ($\ell =E,M,A$)
    for the densities $\rho=0$, $\rho= \rho_0$ and $\rho=2 \rho_0$.
    The magnetic form factors are in natural units
    (magnetic moment: $\mu_N^\ast = G_M^\ast(0) \frac{e}{2 M_N^\ast}$,
    $M_N^\ast$ is the in-medium nucleon mass)~\cite{Medium1}.
    \label{fig-Nucleon1}}}
\end{figure*}

To extend the calculations for higher densities one needs parametrizations 
of the pion decay constant for higher densities.
For that purpose we consider a smooth extension of an in-medium chiral perturbation theory expression for $f_\pi^\ast/f_\pi$~\cite{Kirchbach97}, for densities $\rho > \rho_0$~\cite{Medium1}.
Calculations for the nucleon form factors for $\rho=0$, $\rho_0$, and $2 \rho_0$ are presented in Figure~\ref{fig-Nucleon1}.
From the figures, one can conclude that the nucleon form factors are in general suppressed in the nuclear medium (compared with the free space) and also that the charge and magnetic radii increase in the medium (because the derivatives of the form factors increase in modulus near $Q^2=0$).
The present results are consistent with the suppression of the ratio $G_E/G_M$ in medium measured for the proton at MAMI and JLab~\cite{Dieterich01,Strauch03}, as discussed in Refs.~\cite{Medium2,Medium4}.
The exceptions are the results for the neutron.
In that case we expect an enhancement for the ratio $G_E/G_M$~\cite{Medium2,Medium4}.
Calculations of hyperons can be found in Refs.~\cite{Medium1,Medium2,Medium3}.

We also studied the possibility of using the present formalism (with constant densities) for baryons bound to a nucleus.
In that case we need to calculate the in-medium form factors $G_\ell^\ast(Q^2, \rho_i)$ in terms of the local density $\rho_i$.
The effective in-nucleus form factors
${\overline G_\ell^\ast}$
($\ell =E,M,A$) are defined by~\cite{Lu99a}
\ba
  {\overline G_\ell^\ast} (Q^2) = \frac{4 \pi}{A} \int_0^{\infty} [r^2 \rho(r)] \,
  G_\ell^\ast\large (Q^2, \rho(r) \large) \, dr ,
  \label{eqLocalRho}
  \ea
where $A$ is the nucleus atomic number, and $G_\ell^\ast$ is calculated for the local density $\rho(r)$ at the distance $r$ from the center of the nucleus.
For the discussion, one notices that the weight associated with distance $r$ is determined  by the function $g(r) = r^2 \rho(r)$.

\begin{figure*}[t]
\centerline{
\mbox{
\includegraphics[width=1.9in]{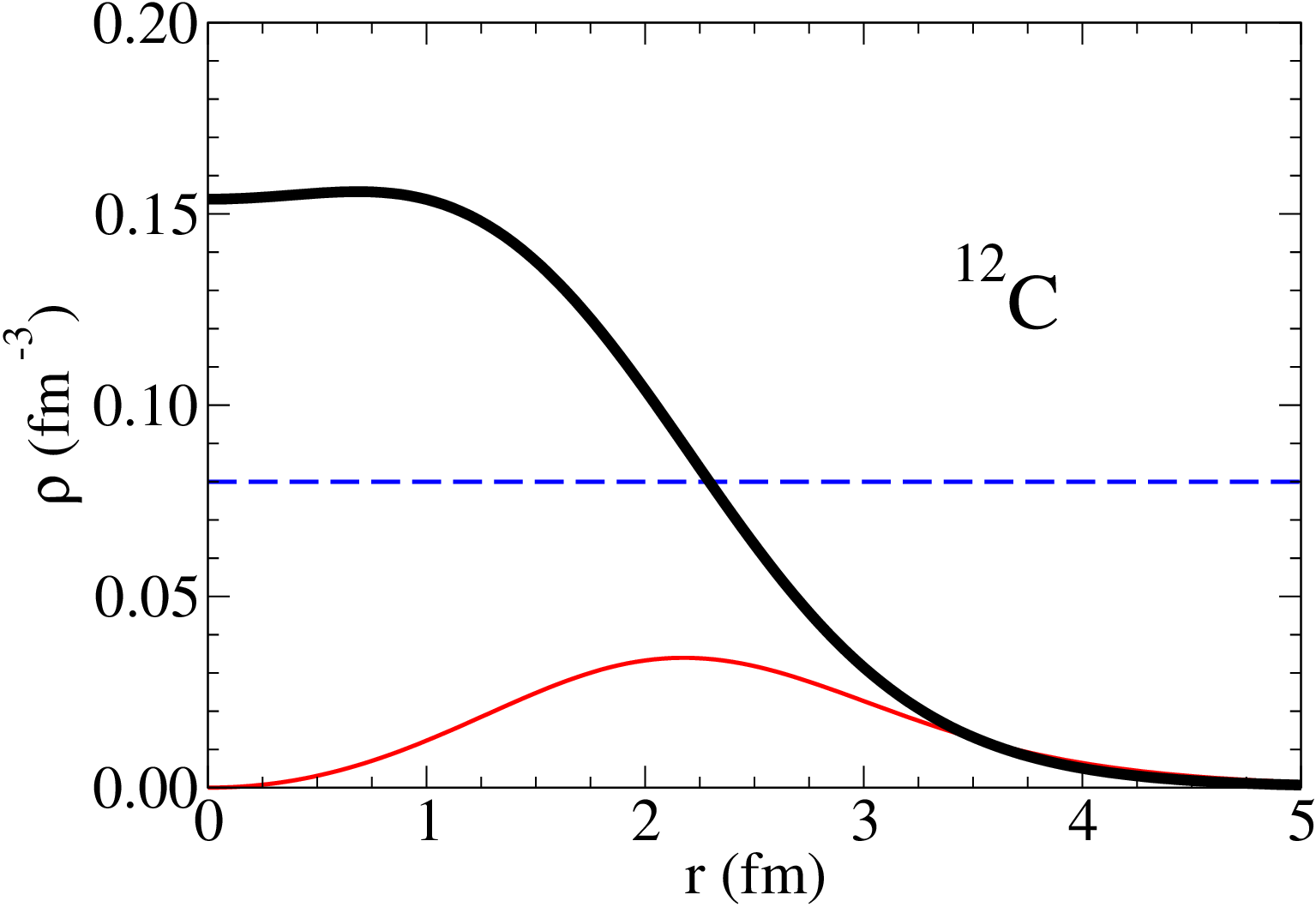} \hspace{.1cm}
\includegraphics[width=1.9in]{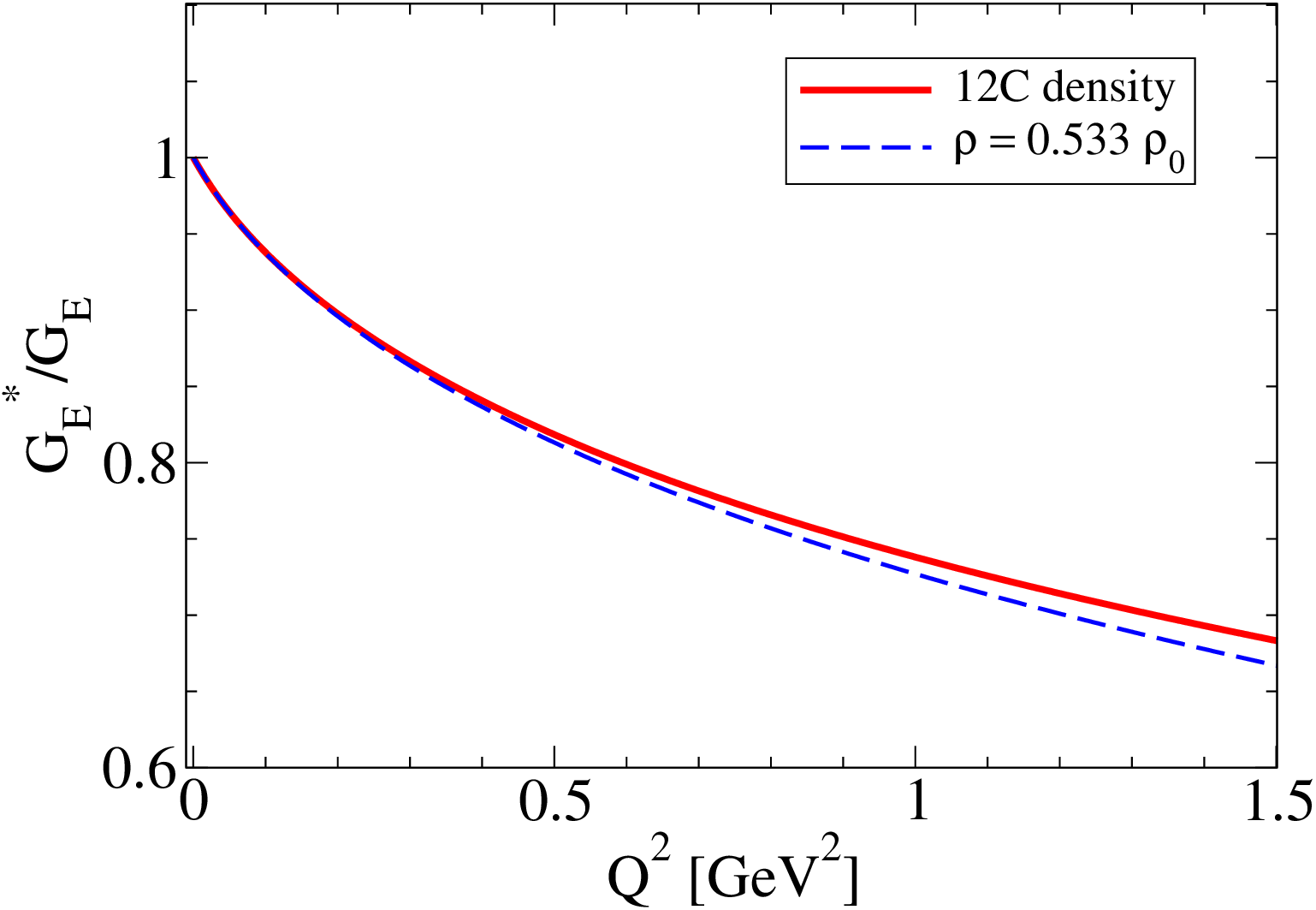} \hspace{.1cm}
\includegraphics[width=1.9in]{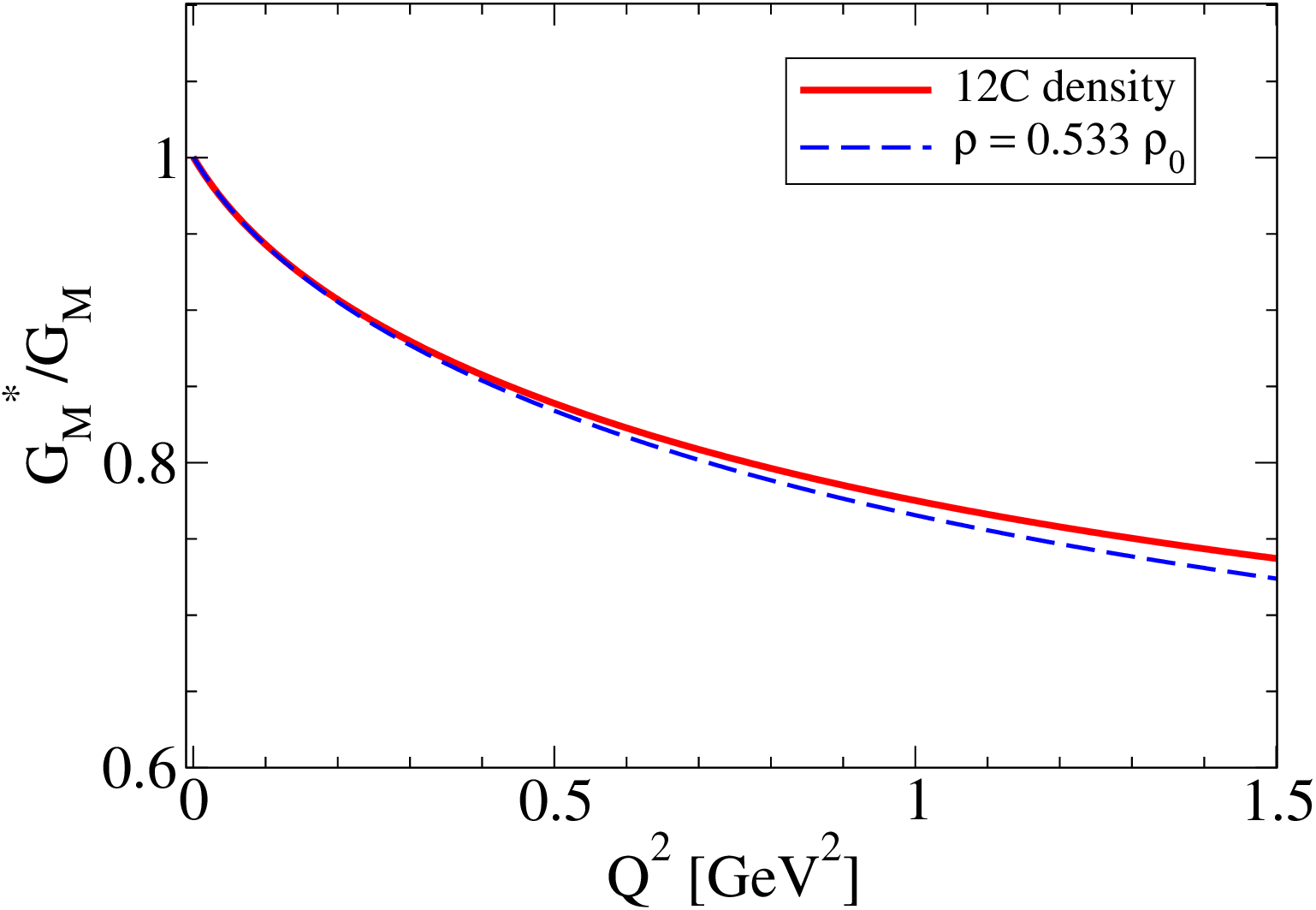}
}}
\caption{\footnotesize{
    {\bf Left:} 
    Nuclear density profile $\rho(r)$ of $^{12}$C.
The dashed line represent the average density $\rho_A$.
{\bf Center:}
   Ratio $G_E^\ast/G_E$ for bound protons.
   {\bf Right:}
   Ratio $G_M^\ast/G_M$ for bound protons.
\label{fig-C12}}}
\end{figure*}

In Figure~\ref{fig-C12}, we present our calculations for a particular nucleus: $^{12}$C.
The nuclear density profile function, determined by the QMC model~\cite{QMCEMFFMedium5}, is presented in the left panel (thick solid line).
The dashed line represents the average density, and the thin solid line the function $g(r)$ in relative units.
In the figure, one can notice that the function $g(r)$ has a maximum in the region associated with the surface of the nucleus (central region).
The maximum of $g(r)$ appears then at the point where the function $\rho(r)$ is very close to the average density $\rho_A$.

The ratios between in-medium form factors and free space form factors for the electric and magnetic form factors of the proton are presented in the center and right panels of Figure~\ref{fig-C12}.
From the figures, we can conclude that the calculation that uses the nuclear density profile function (solid line), and the calculation using the average density are very close below $Q^2=0.2$ GeV$^2$.
  Only above this point does one start to notice differences (on the order of a few percent).

  These results justify the use of calculations of form factors of bound nucleons based on the average density at low-$Q^2$.
The approximation is then justified in studies of bremsstrahlung of nuclei~\cite{Maydanyuk26}.
In progress are calculations of the proton double ratios $(G_E^\ast/G_M^\ast)/(G_E/G_M)$ for the nucleus $^{12}$C, $^{16}$O and $^{40}$Ca~\cite{InPreparation}.
Our predictions can be compared with data extracted from the polarization transfer measurements at MAMI~\cite{Izraeli18a,Kolar23a}, and with future experiments.

\vspace{-1.3ex}

\section{Outlook and conclusions}

Calculations of the electroweak form factors of baryon in a nuclear medium are fundamental to study interactions in dense nuclear matter.
We calculate the electroweak form factors of the octet baryon for different constant densities.
We conclude that the nuclear medium modifies the baryon properties differently (quenched or enhanced), according to the mass and flavor content of the baryons, and that the impact of the medium increases, in general, with the density.

We also studied the form factor modifications of nucleons in finite nuclei.
We conclude that the effect of the shape of the density function is small for low $Q^2$.
In progress are calculations for higher densities (up to $\rho=3 \rho_0$), that can be used for studies with astrophysical implications.


\vspace{-1.4ex}

\acknowledgments

G.R.~and M.-K.C.~were supported by the National
Research Foundation of Korea (Grant  No.~RS-2021-NR060129).
K.T.~was supported by CNPq, Brazil, Processes 
No.~304199/2022-2, FAPESP Process No.~2023/07313-6,
and by Instituto Nacional de Ci\^{e}ncia e Tecnologia -- Nuclear Physics and Applications
(INCT-FNA), Brazil, Process No.~464898/2014-5.





\end{document}